\documentclass{ismdproc}

\begin{document}
%------------------------------------
\title{Boost-invariant one-tube model for two-particle correlation}

\author{{\slshape Rone Andrade$^1$\footnote{Speaker}, Fr\'ed\'erique Grassi$^1$,
    Yogiro Hama$^1$, Wei-Liang Qian$^2$}\\[1ex]
$^1$Instituto de F\'isica, Universidade de S\~ao Paulo, C.P. 66318, 05315-970 S\~ao Paulo-SP, Brazil\\
$^2$Departamento de F\'{\i}sica, Universidade Federal de Ouro Preto, Ouro Preto-MG, Brazil}

% please do not modify the following 5 lines
\contribID{xy}  % will be entered by the editors
\confID{yz}
\acronym{ISMD2010}
\doi            % will be entered by the editors

\maketitle

\begin{abstract}
We show the in-plane/out-of-plane effect, in the two-particle correlation function, computed with the NexSPheRIO code,
in Au+Au collisions at 200AGeV. In order to clarify the origin of the effect, a simplified model, which consists
of a peripheral high-energy-density tube in a smooth background with longitudinal boost invariance, is applied.
\end{abstract}

\section{Introduction}
\label{sec:introduction}

\begin{figure}[h!b]
\centerline{\includegraphics[width=0.36\textwidth]{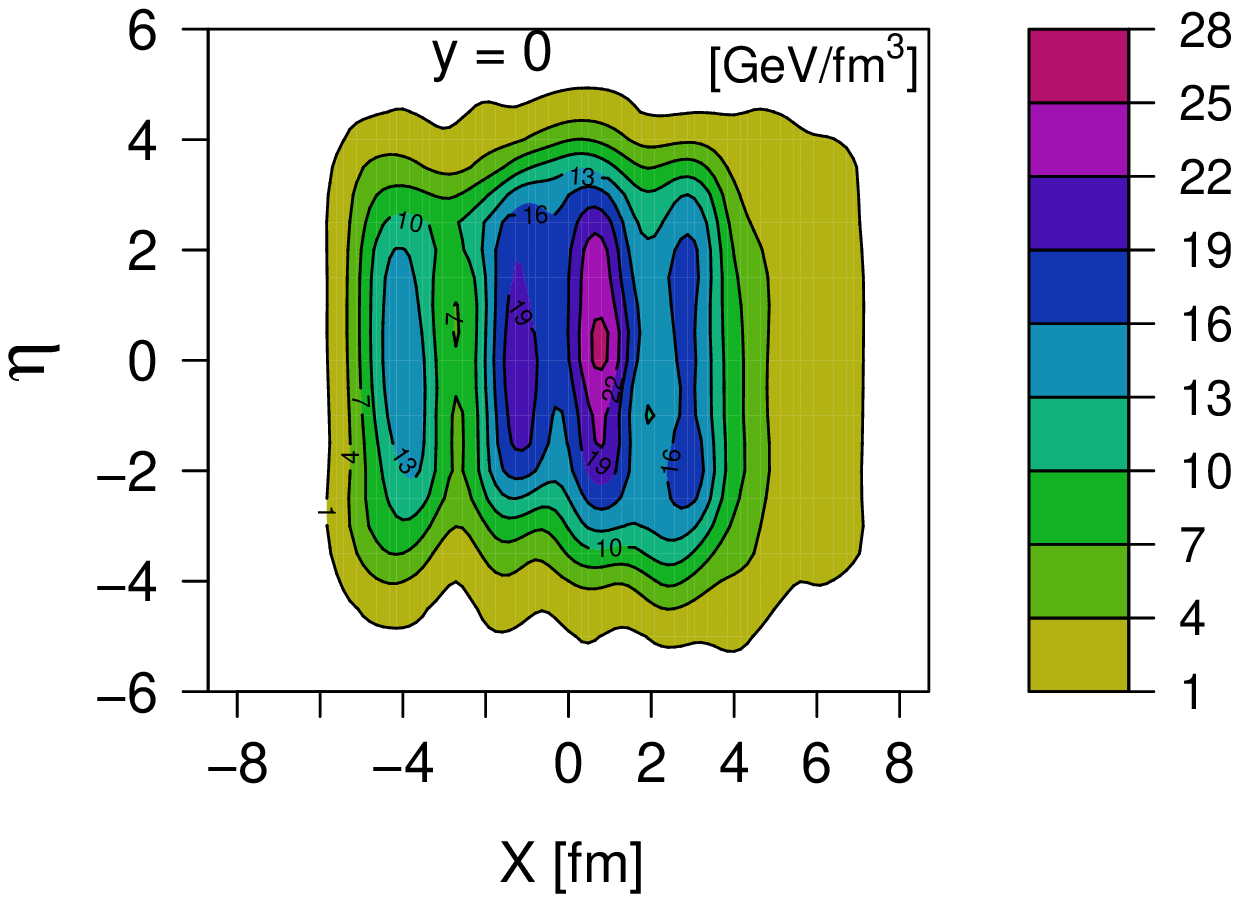}
            \includegraphics[width=0.36\textwidth]{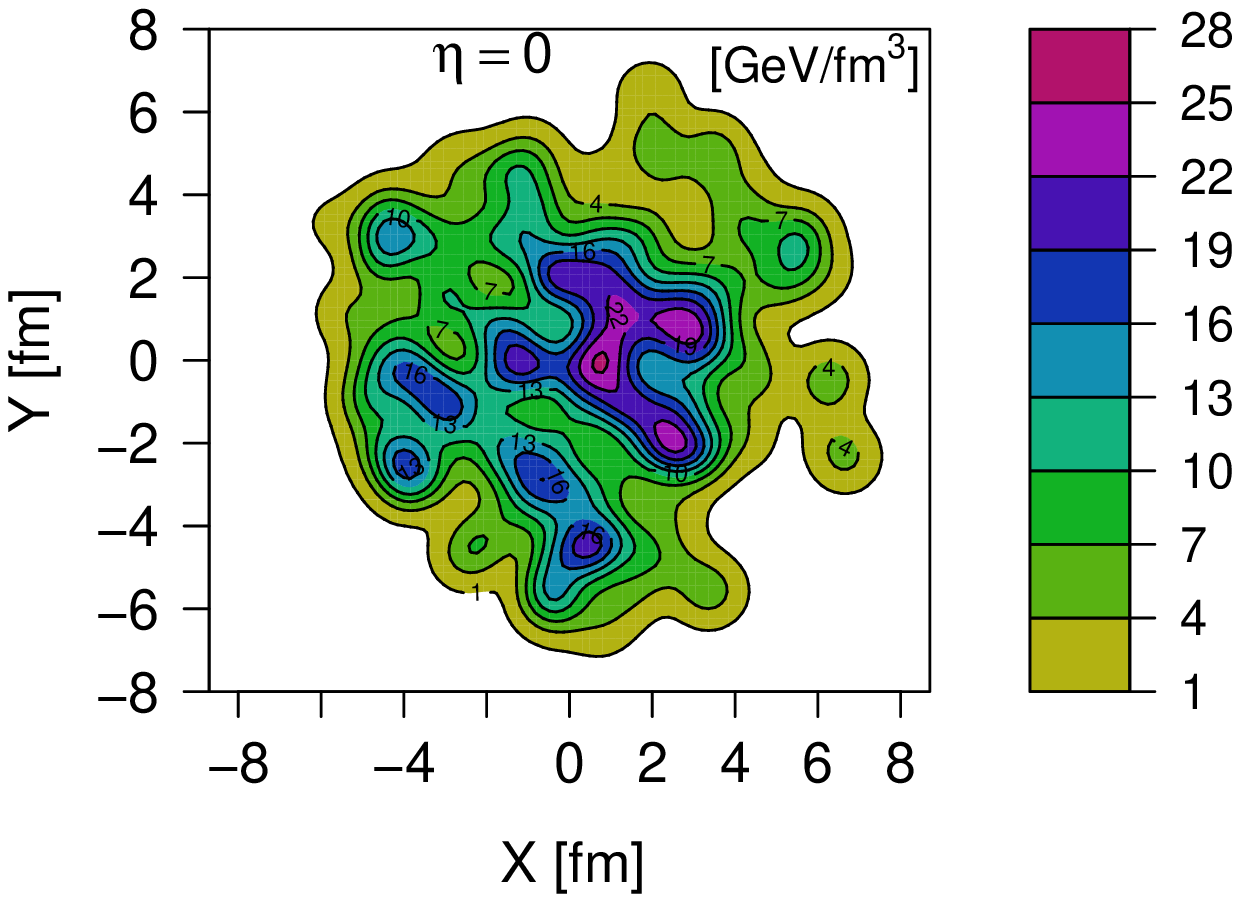}}
\caption{initial energy density distribution of a random Nexus event
(central Au+Au collision at 200AGeV).}\label{Fig:nexus.ic}
\end{figure}

In the previous papers \cite{Takahashi:2009na,Hama:2009vu,Andrade:2009em,Andrade:2010xy}, we have shown that the
two-particle correlation computed with NexSPheRIO is in good agreement with the main
characteristics of the Au+Au data at RHIC \cite{ridge1,ridge1s,ridge2,dhump1,phobos,phoboss,Feng:2008an}.
The NexSPheRIO code is a junction of the event generator Nexus \cite{Drescher:2000ec} and the hydrodynamic code
SPheRIO \cite{Hama:2004rr}. The ridge structure, in this model, is related to the tubular structures that characterize
the Nexus initial conditions. In Fig.\ref{Fig:nexus.ic}, an example of initial energy density distribution is shown.
In particular, the existence of peripheral tubes is the crucial ingredient to produce ridges, for instance the tube
at $x \sim -4fm$ (see the left-hand figure). For a general discussion of our method, see also the talk presented by
Y.Hama in this Symposium \cite{hama:ismd2010}.

An interesting effect that has been experimentally studied is the dependence of the two-particle correlation function
on the azimuthal angle of the trigger particle $\phi_{\operatorname{s}}$ with respect to the event plane \cite{Feng:2008an}.
In a mid-central window, the away-side structure in $\Delta \phi$ is a peak at $\pi$, if the trigger is close to the event
plane, and it is split into two peaks, as $\phi_{\operatorname{s}}$ goes closer to $\pi/2$. This is the so called
in-plane/out-of-plane effect. In the next Section, we show the in-plane/out-of-plane correlation computed with the NexSPheRIO
code, in Au+Au collisions at 200AGeV. In Sec.\ref{sec:one-tube_model}, the one-tube model is applied to understand the role
of peripheral tubes in creating the effect. Our conclusions are summarized in Sec.\ref{sec:conclusion}.

\section{NexSPheRIO results}
\label{sec:nexspherio_results}

\begin{figure}[h!b]
\centerline{\includegraphics[width=0.48\textwidth]{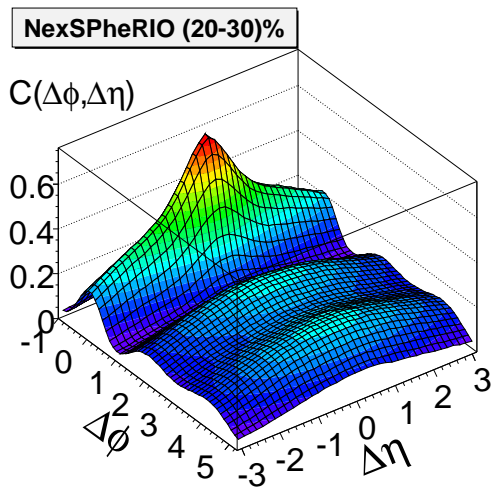}}
\caption{two-particle correlation function $C=\left(1/N_{tr} \right) dN/\left( d\Delta \phi d\Delta \eta \right)$
computed with NexSPheRIO, in the (20-30)$\%$ centrality window, for Au+Au collisions at 200AGeV. The associated particles
were chosen with $p_t > 1.0$GeV and the triggers with $p_t > 2.5$GeV.}\label{Fig:nexspherio.corr}
\end{figure}

In Fig.\ref{Fig:nexspherio.corr}, the two-particle correlation function computed with NexSPheRIO is shown, for Au+Au
collisions at 200AGeV. Both the near-side ridge structure and the away-side double hump structure are reproduced. The
shape of the latter is elongated in the $\eta$ direction as well, showing as a double ridge.
The events in the (20-30)$\%$ centrality window were chosen according to the number of participant nucleons,
from 140 to 195. The mixed event method was applied to remove the correlation due to both the shape of the $\eta$
distribution and the elliptic flow. In our version of this technique, each event is rotated, in order
to align the reaction planes, before computing the mixed event term of the correlation.

\begin{figure}[h!t]
\centerline{\includegraphics[width=0.33\textwidth]{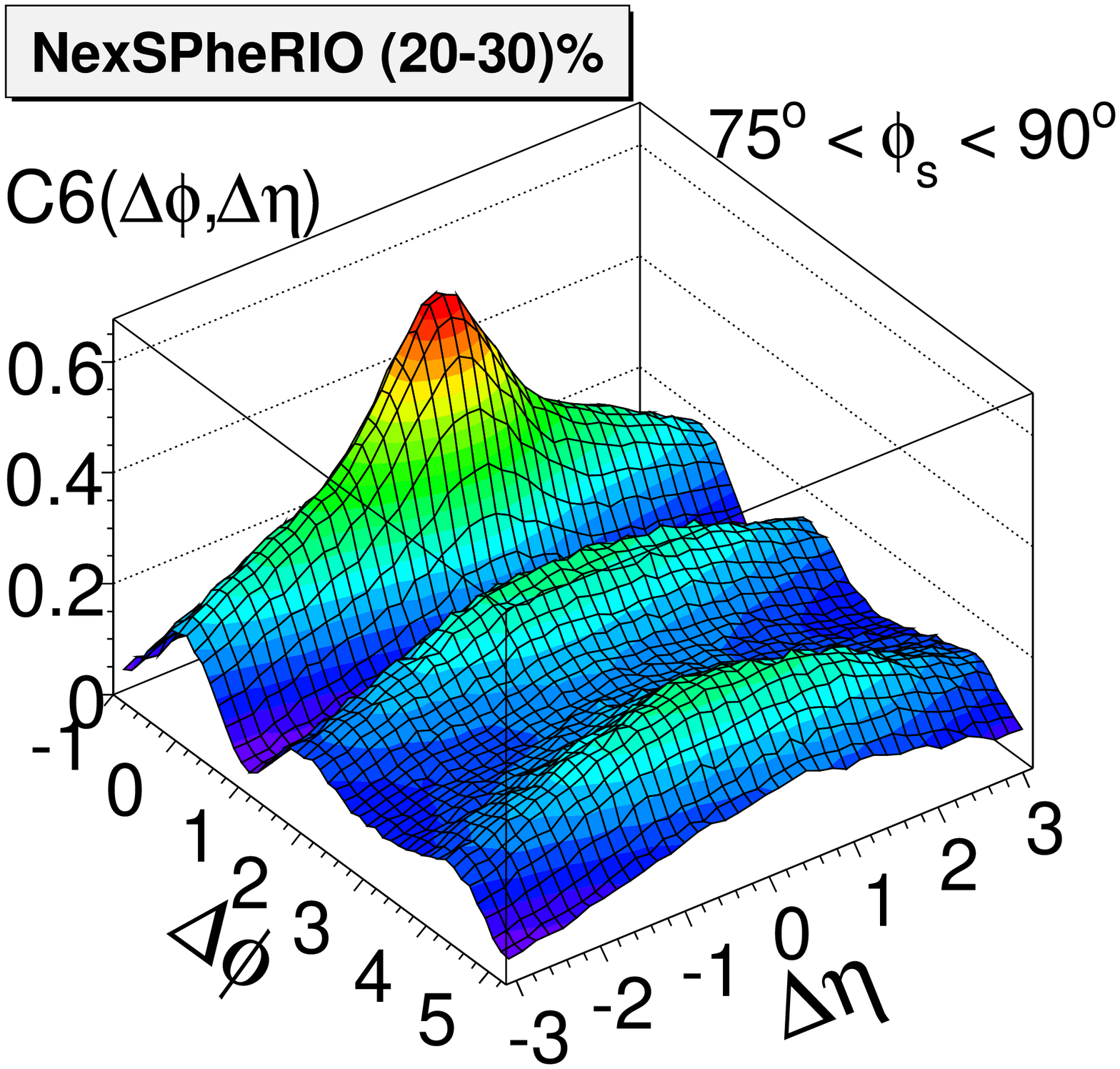}
            \includegraphics[width=0.33\textwidth]{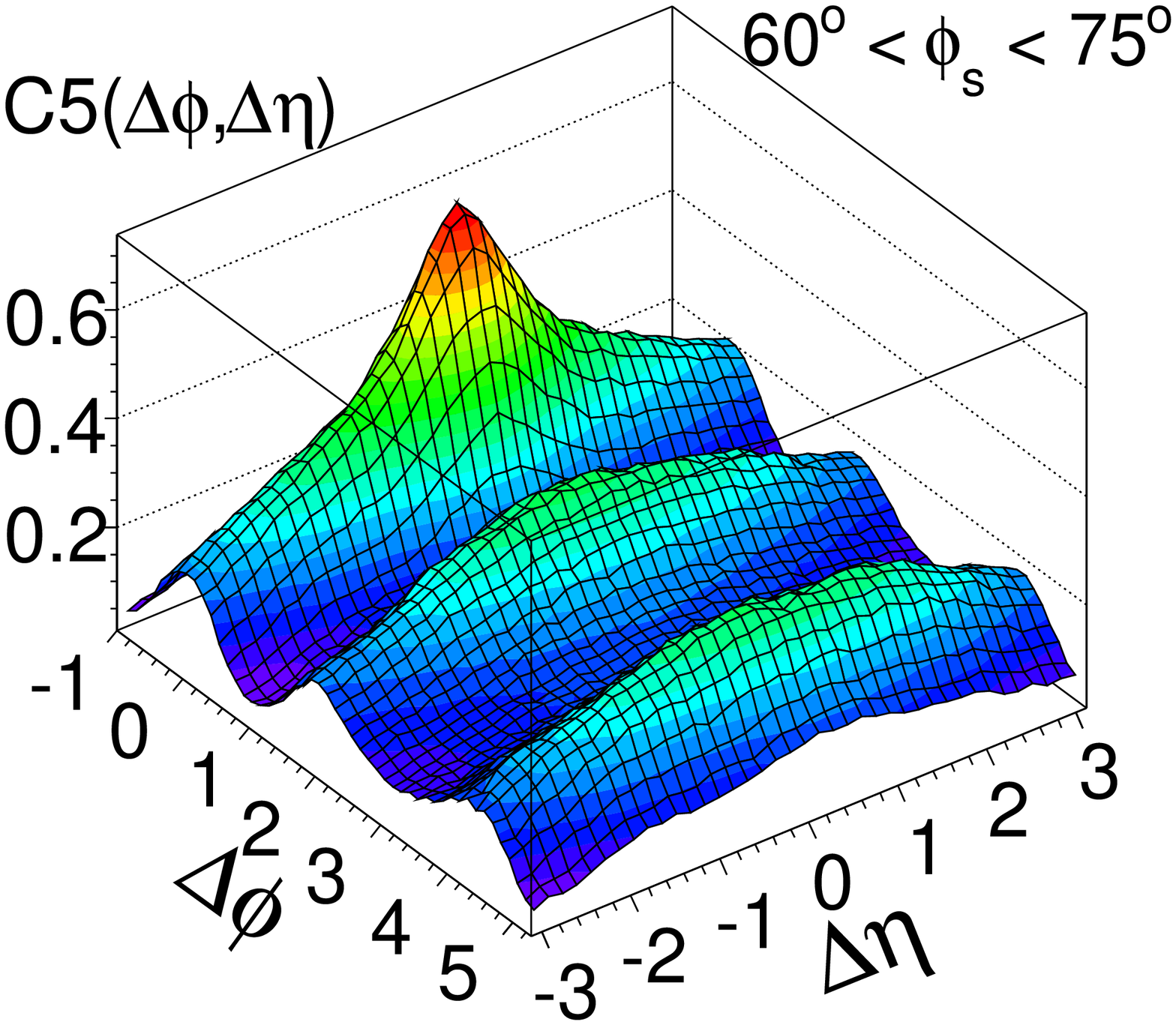}
            \includegraphics[width=0.33\textwidth]{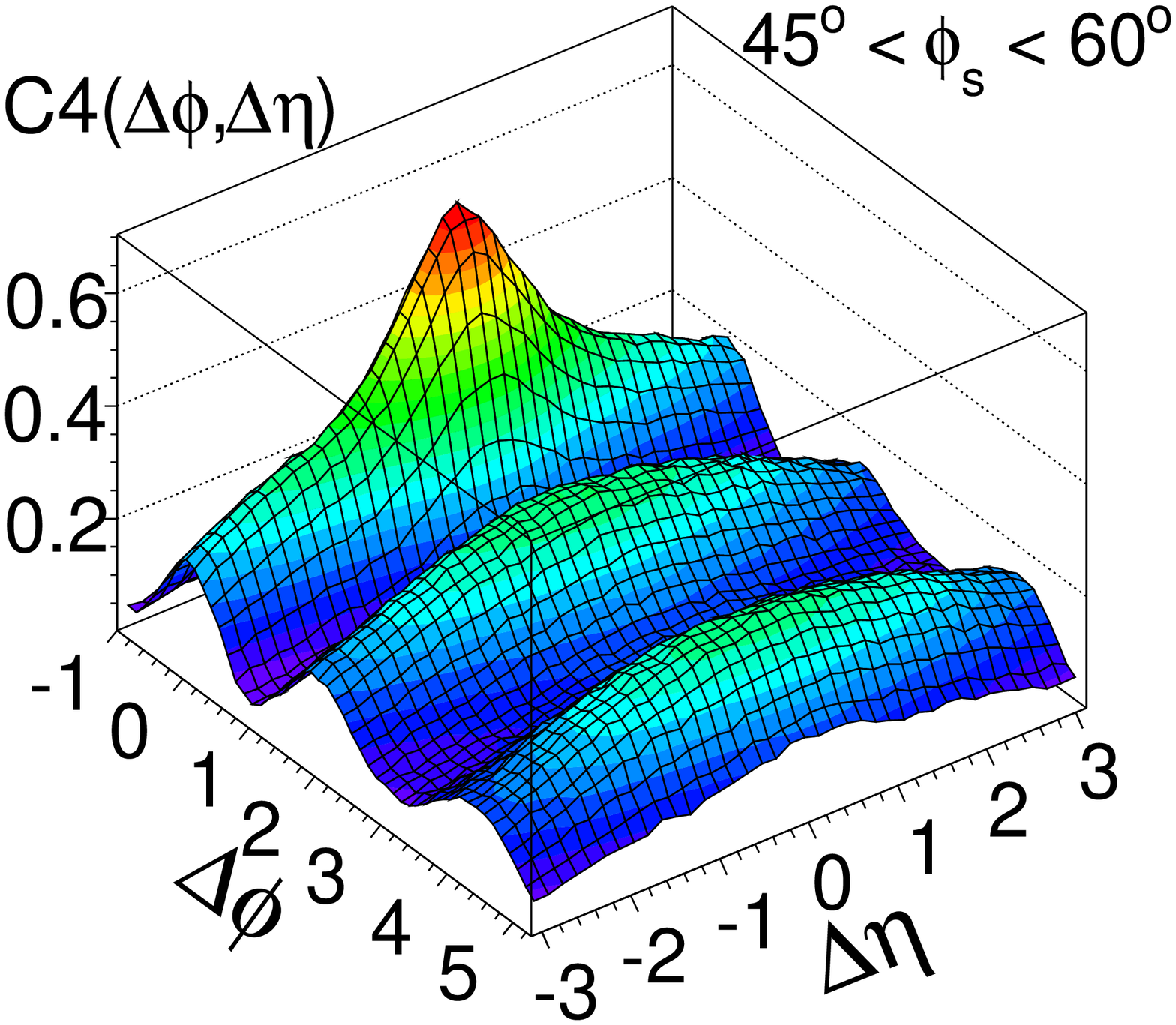}}
\centerline{\includegraphics[width=0.33\textwidth]{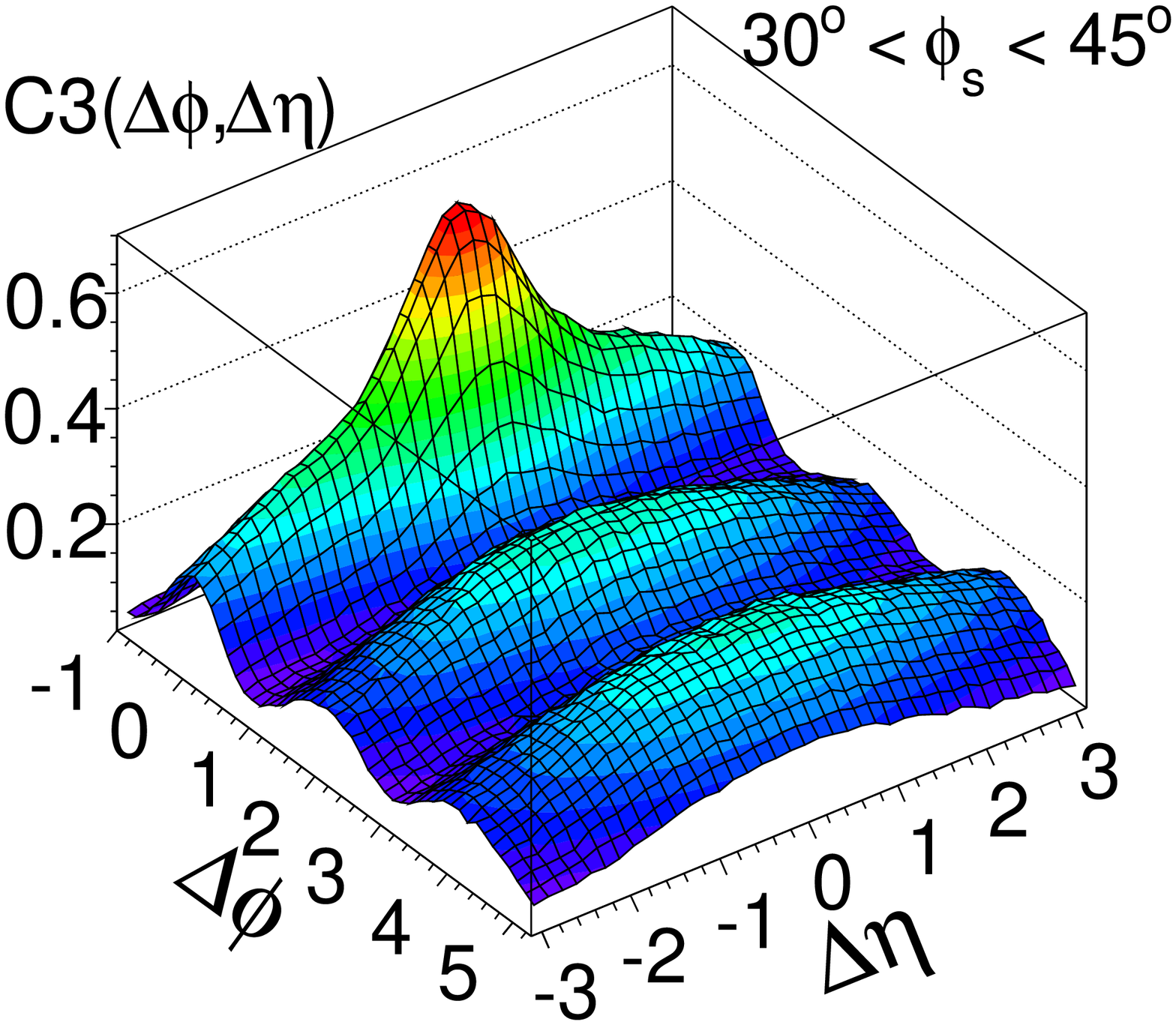}
            \includegraphics[width=0.33\textwidth]{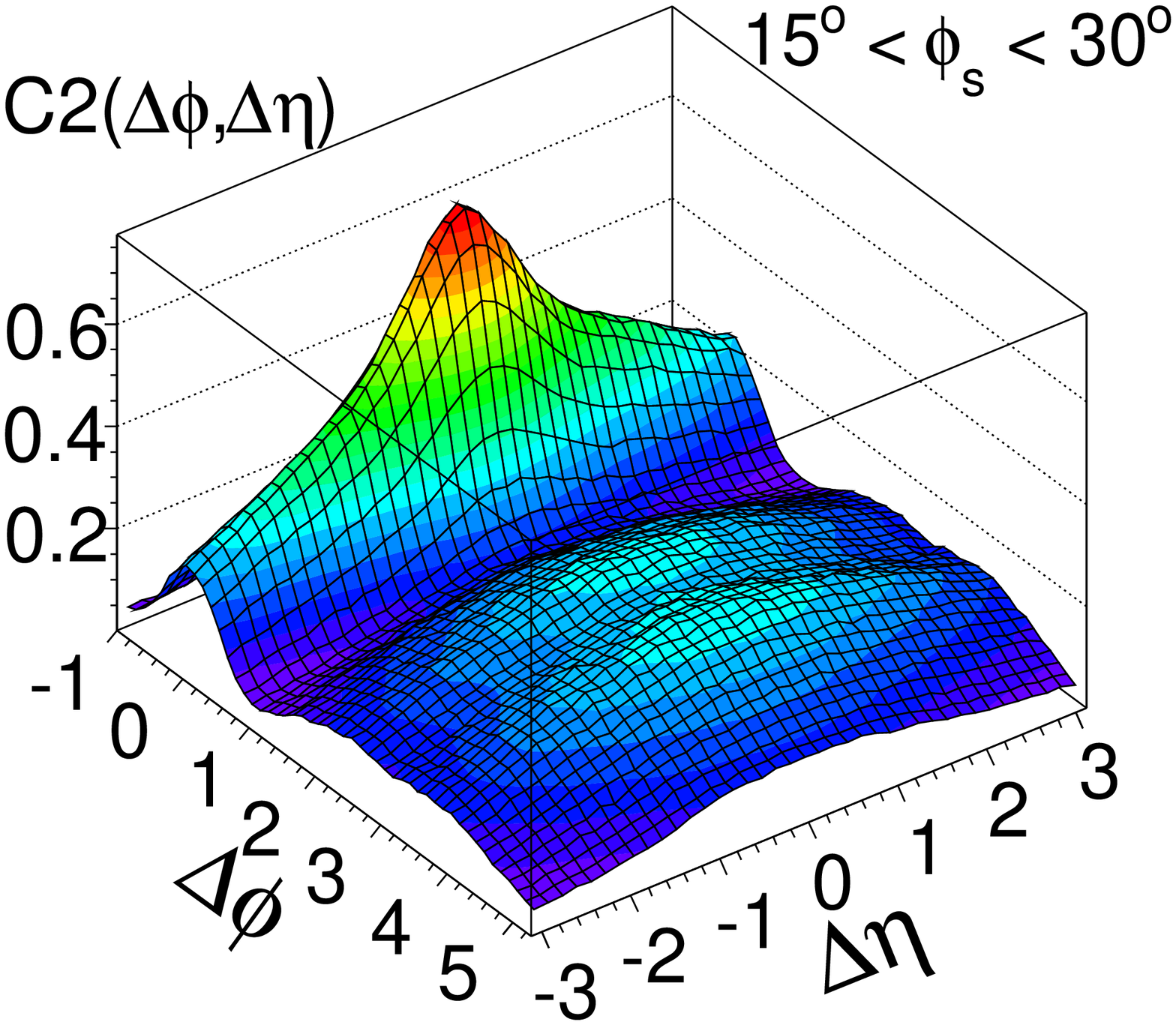}
            \includegraphics[width=0.33\textwidth]{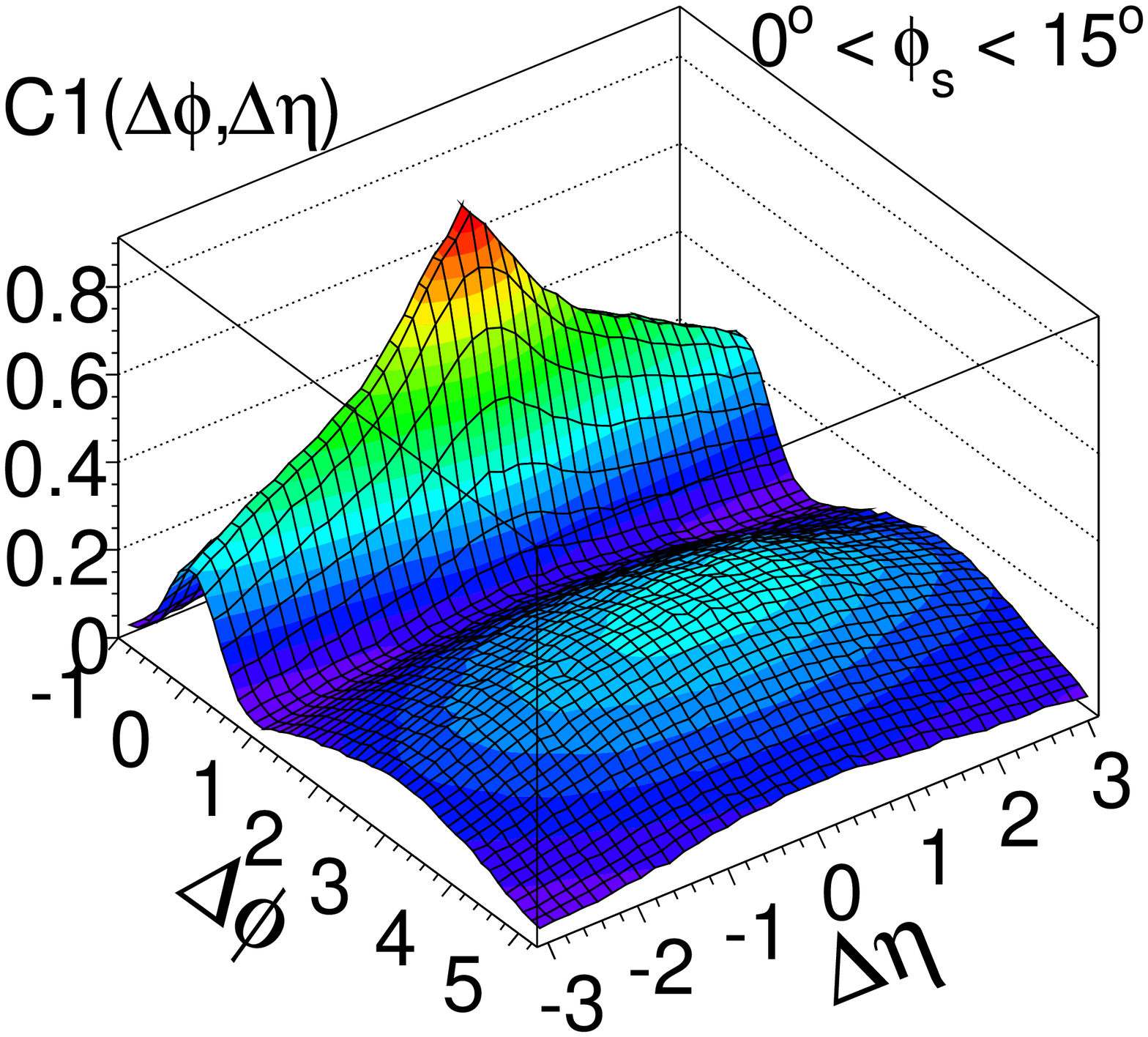}}
\caption{in-plane/out-of-plane correlation computed with NexSPheRIO, in the (20-30)$\%$ centrality window,
for Au+Au collisions at 200AGeV. From top to bottom and left to right: C6 ($75^{o}<\phi_{s}<90^{o}$),
C5 ($60^{o}<\phi_{s}<75^{o}$), C4 ($45^{o}<\phi_{s}<60^{o}$), C3 ($30^{o}<\phi_{s}<45^{o}$),
C2 ($15^{o}<\phi_{s}<30^{o}$), C1 ($0^{o}<\phi_{s}<15^{o}$). $\phi_{s}$ is the azimuthal angular difference
between the trigger momentum and the event plane.}\label{Fig:nexspherio.corr.phis}
\end{figure}

In Fig.\ref{Fig:nexspherio.corr.phis}, our results on in-plane/out-of-plane correlation are shown, within the same
centrality window. The away-side structure evolves from a double to a single ridge as the
azimuthal angle of the trigger goes from $90^{\operatorname{o}}$ to $0^{\operatorname{o}}$ with respect
to the event plane, just as observed in data \cite{Feng:2008an}.

\section{One-tube model}
\label{sec:one-tube_model}

In previous papers \cite{Hama:2009vu,Andrade:2009em,Andrade:2010xy}, we have introduced
the one-tube model, which is a suitable tool for studying the hydrodynamic expansion in the neighborhood of a peripheral
high-energy-density tube. In this model, a complex bulk of matter generated by Nexus is replaced by a smooth profile of
energy density (background), and a tube (or hot spot) is placed at some position along a contour curve close to the border.
For the longitudinal expansion, boost invariance is assumed. The aim of this Section is to try to understand the NexSPheRIO
results on in-plane/out-of-plane correlation (given in previous Section), by using this simplified model.

\begin{figure}[h!b]
\centerline{\includegraphics[width=0.47\textwidth]{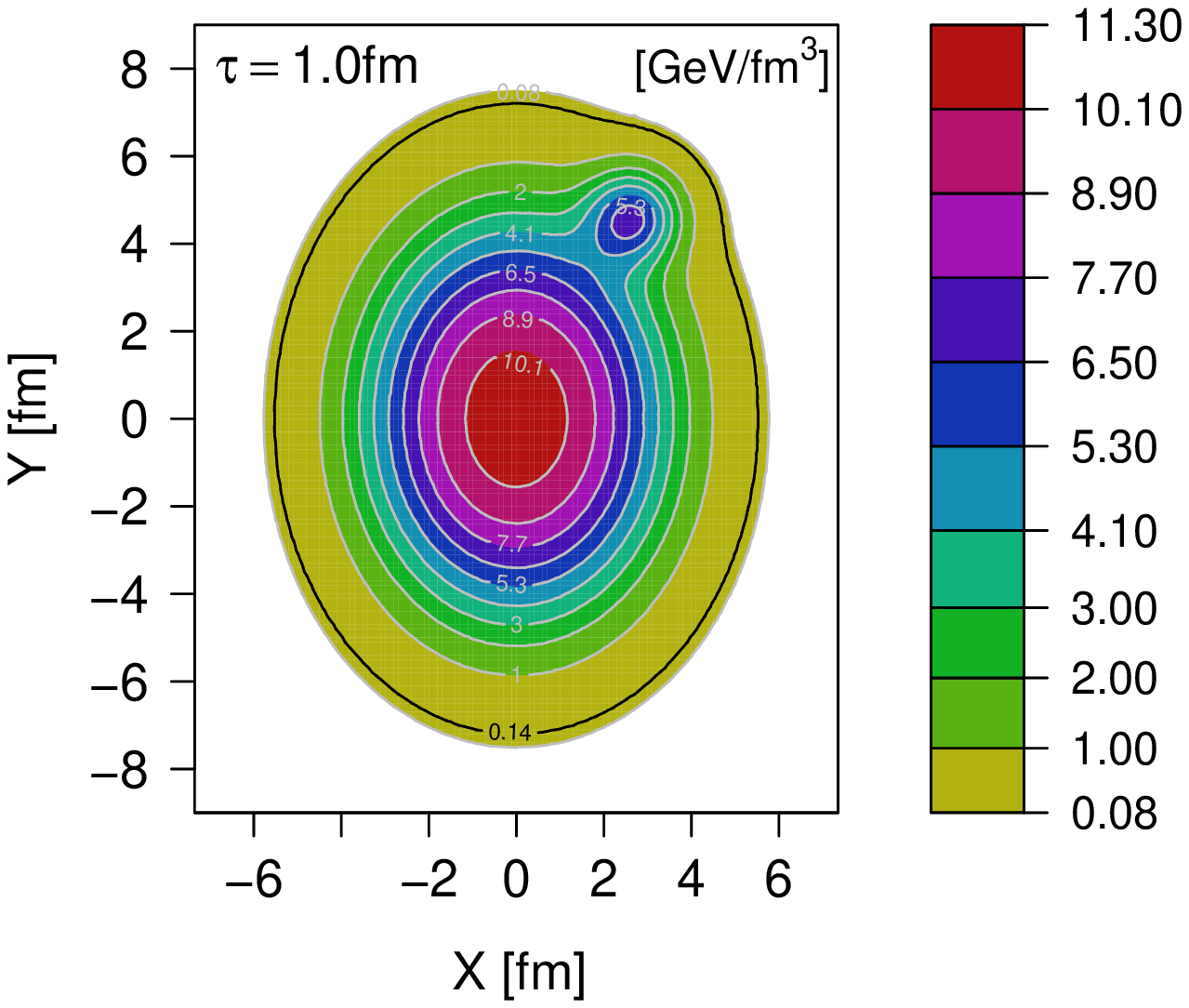}
            \includegraphics[width=0.46\textwidth]{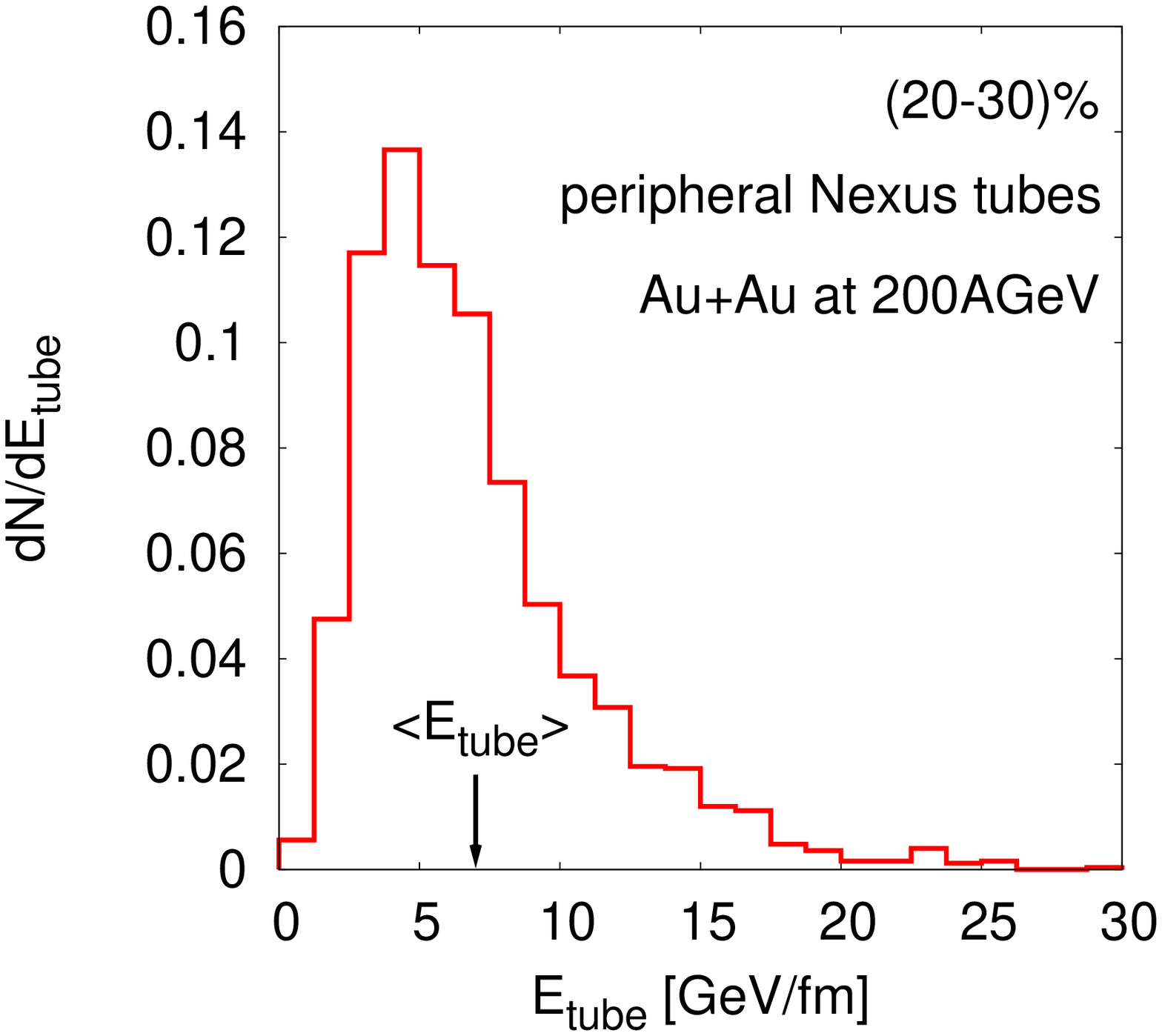}}
\caption{left: energy density distribution for the one-tube event with $\phi_{\operatorname{tube}}=60^{\operatorname{o}}$.
The tube is placed on the contour curve $\operatorname{\epsilon_{back}} \sim 1$GeV/fm$^3$.
Right: energy-content distribution ($\operatorname{E_{tube}} \propto \epsilon \Delta r^2$) of
peripheral Nexus tubes, in the (20-30)$\%$ centrality window. The tubes used in this plot
were selected, in each event, from the transverse area limited by the ellipses
$\operatorname{r_{1}}=5.18/\operatorname{R_{n}}$ and 
$\operatorname{r_{2}}=5.81/\operatorname{R_{n}}$, where
$\operatorname{R_{n}}=\sqrt{1.74 \cos\left(\phi-\Psi_{n} \right)^2 + \sin\left(\phi-\Psi_{n} \right)^2}$
and $\Psi_{n}$ is the event plane angle of the nth event.}\label{Fig:one-tube.ed}
\end{figure}

In Fig.\ref{Fig:one-tube.ed} (left), an example of initial energy density distribution in the one-tube model is shown.
The background is the average of Nexus initial energy density in the (20-30)$\%$ centrality window, for Au+Au collisions
at 200AGeV, and the peripheral tube is placed at some point along the the contour curve
$\operatorname{\epsilon_{bkgd}}  \sim 1$GeV/fm$^3$. The tube has an energy content $\operatorname{E_{tube}} \sim 7$ Gev/fm,
which is compatible with the typical energy content of a peripheral Nexus tube, in the same centrality window
(see the right-hand figure). The energy content, or energy per unit of length
($\operatorname{E_{tube}} \propto \epsilon \Delta r^2$), as we have shown \cite{Andrade:2010xy}, is a suitable parameter
to characterize tubes, once the relevant factor in this model is the capacity of expansion of these objects.
(For instance, if the energy per unit of length of the tube is kept constant, while its radius
is varied, the overall shape of the two-particle correlation is preserved). The parametrization of the initial energy density,
for the one-tube event with $\phi_{\operatorname{tube}}=60^{\operatorname{o}}$, is given by:

\begin{equation}
\operatorname{\epsilon_{bkgd}} + \operatorname{\epsilon_{tube}} = 10.8 \operatorname{exp} \left[ - \left( \frac{1.74x^2 + y^2}{\left( 4.35\right)^2}  \right)^{\frac{3}{2}} \right]+
6.2 \operatorname{exp} \left[ \frac{-\left( x-2.7  \right)^2 - \left( y-4.6 \right)^2}{\left( 1.0\right)^2}  \right]
\end{equation}

\noindent
and both the baryon density and the initial transverse velocity are assumed null.

\begin{figure}[h!b]
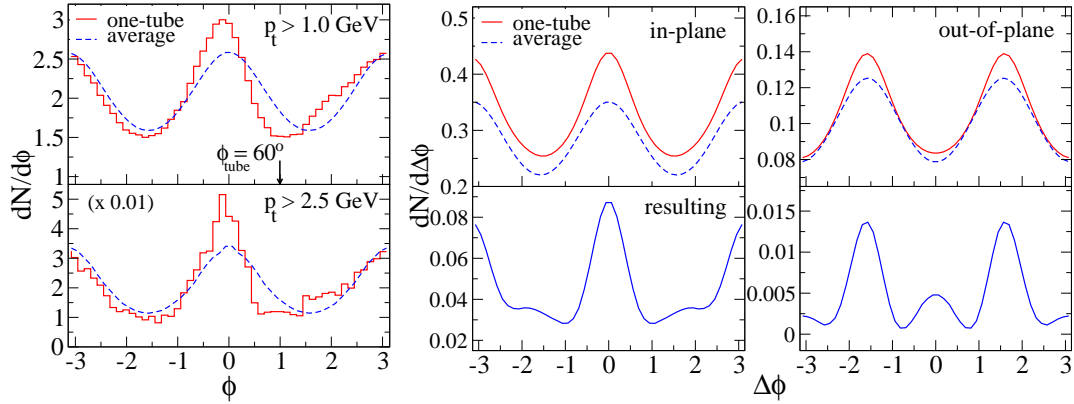

\centerline{\includegraphics[width=0.35\textwidth]{andrade_rone.fig13.eps}\hspace{1mm}
            \includegraphics[width=0.61\textwidth]{andrade_rone.fig14.eps}}
\caption{left: azimuthal distribution of the associated particles (top) and triggers (bottom). The solid lines refer to the
one-tube event with $\phi_{\operatorname{tube}}=60^{\operatorname{o}}$ and the dashed lines to the average angular
distribution. Right: correspondig two-particle correlation functions (upper panels) and the resulting correlation
(lower panels), for in-plane and out-of-plane triggers.}\label{Fig:corr.60}
\end{figure}

In Fig.\ref{Fig:corr.60} (left), the azimuthal distribution of the associated particles (top) and triggers (bottom)
are shown. The solid lines refer to the one-tube event with $\phi_{\operatorname{tube}}=60^{\operatorname{o}}$
and the dashed lines to the average angular distribution. In the latter, the average is performed over a
set of one-tube events with the tube randomly placed along the same contour curve.
In both plots the behavior is similar: there is a lack of particles at the angular position of the tube, $\phi \sim \pi /3$,
and an excess at $\phi \sim 0$ and $\phi \sim \pi /2$, in comparison with the average angular distribution. The shadowing
effect produced by the peripheral tube here is similar to what we have found in the central collisions version of this model
\cite{Hama:2009vu,Andrade:2009em,Andrade:2010xy}. The two-particle correlation functions are shown in the same figure (right),
for in-plane and out-of-plane triggers, respectively. The upper panels refer to the one-tube event with
$\phi_{\operatorname{tube}}=60^{\operatorname{o}}$ (solid lines) and the average correlation (dashed lines). The difference
between both functions is shown in the lower panels (resulting correlation). The symmetrical one-tube events are included:
$120^{\operatorname{o}}$, $240^{\operatorname{o}}$ and $300^{\operatorname{o}}$, which turns the plots symmetrical with
respect to the origin.

In the in-plane case, in which the azimuthal angle of the trigger goes from $-15^{\operatorname{o}}$ to
$15^{\operatorname{o}}$ (and from $165^{\operatorname{o}}$ to $195^{\operatorname{o}}$), the two-particle correlation shows
a peak at 0 and at $\pi$ as a consequence of the excess of particles at $\phi \sim 0$. The excess of
associated particles at $\phi \sim \pi/2$ gives rise to the local maximal point at $\Delta \phi \sim \pi/2$.
In the out-of-plane case, the excess of associated particles at $\phi \sim 0$ gives rise to a peak at $\pi/2$ and at $-\pi/2$,
once the azimuthal angle of the trigger goes from $75^{\operatorname{o}}$ to $105^{\operatorname{o}}$
(and from $255^{\operatorname{o}}$ to $285^{\operatorname{o}}$). Finally, there is a central peak (lower) due to the excess of
particles at $\phi \sim \pi/2$. This result shows that the near-side and away-side structure, the latter with one or two
peaks, can be understood in terms of the hydrodynamic expansion of the matter in the neighborhood of a peripheral
high-energy-density tube.

\begin{figure}[h!t]
\centerline{\includegraphics[width=0.47\textwidth]{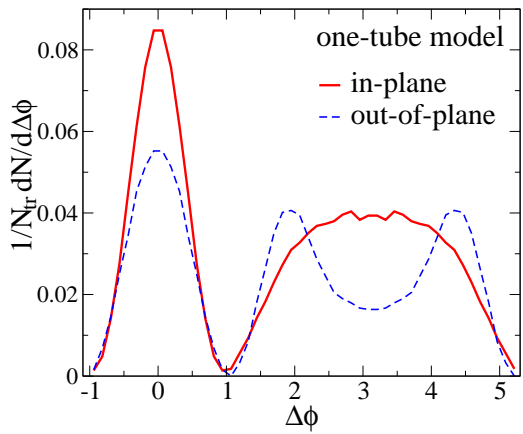}}
\caption{the final in-plane/out-of-plane correlation in the one-tube model, normalized by the number of
triggers.}\label{Fig:corr.in-ou}
\end{figure}

By moving the tube position along the contour curve $\operatorname{\epsilon_{bkgd}}  \sim 1$GeV/fm$^3$, the position
of peaks and valleys change. (For instance, a configuration with $\phi_{\operatorname{tube}}=0^{\operatorname{o}}$ produces
a valley at $\phi = 0$, in the azimuthal distribution). However, when the two-particle correlation is integrated over tube
position in order to obtain the final correlation for in-plane (or out-of-plane) triggers, certainly those configurations
close to the example we chose ($\phi_{\operatorname{tube}}=60^{\operatorname{o}}$) dominate, because they produce an excess
of particles, in comparison with the averaged azimuthal distribution, at $\phi \sim 0$ (in-plane) and $\phi \sim \pi/2$
(out-of-plane). So, the results obtained with this particular configuration can be used to understand the overall shape of
the integrated correlation. In Fig. \ref{Fig:corr.in-ou}, such final in-plane/out-of-plane correlations are shown,
normalized by the number of triggers.

\section{Conclusion}
\label{sec:conclusion}
We discussed the in-plane/out-of-plane effect in the NexSPheRIO scenario, for Au+Au collisions at 200AGeV. It was
shown that the qualitative behavior of the correlation function is consistent with the data at RHIC. In addition,
a simplified model (one-tube model) was applied to clarify the origin of the effect. It was observed that the overall
shape of the in-plane/out-of-plane correlations are reproduced by the model as a consequence of the hydrodynamic
expansion of the matter in the neighborhood of a peripheral high-energy-density tube.

\section{Acknowledgments}
\label{sec:acknowledgments}

We acknowledge funding from CNPq and FAPESP. 

% ****************************************************************************
% BIBLIOGRAPHY AREA
% ****************************************************************************

% please do not change the following line
\begin{footnotesize}

% please do not change the following line
\end{footnotesize}

% ****************************************************************************
% END OF BIBLIOGRAPHY AREA
% ****************************************************************************


\begin{thebibliography}{99}

%------- replace following references ;-)

\bibitem{Takahashi:2009na}
  J.~Takahashi {\it et al.},
  %``Topology studies of hydrodynamics using two particle correlation
  %analysis,''
  Phys.\ Rev.\ Lett.\  {\bf 103}, 242301 (2009)
  [arXiv:0902.4870 [nucl-th]].
  %%CITATION = PRLTA,103,242301;%%

\bibitem{Hama:2009vu}
  Y.~Hama, R.~P.~G.~Andrade, F.~Grassi and W.~L.~Qian,
  %``Trying to understand the ridge effect in hydrodynamic model,''
  Nonlin.\ Phenom.\ Complex Syst.\  {\bf 12}, 466 (2009)
  [arXiv:0911.0811 [hep-ph]].
  %%CITATION = NPCSC,12,466;%%

\bibitem{Andrade:2009em}
  R.~Andrade, F.~Grassi, Y.~Hama and W.~L.~Qian,
  %``A closer look at the influence of tubular initial conditions on
  %two-particle correlations,''
  J.\ Phys.\ G {\bf 37}, 094043 (2010)
  [arXiv:0912.0703 [nucl-th]].
  %%CITATION = JPHGB,G37,094043;%%

\bibitem{Andrade:2010xy}
  R.~P.~G.~Andrade, F.~Grassi, Y.~Hama and W.~L.~Qian,
  %``Influence of tubular initial conditions on two-particle correlations in
  %relativistic nuclear collisions,''
  arXiv:1008.4612 [nucl-th].
  %%CITATION = ARXIV:1008.4612;%%

\bibitem{ridge1} 
   J. Putschke (for the STAR collaboration), 
   Nucl. Phys. A{\bf 783} (2007) 507. 

\bibitem{ridge1s}
   J. Putschke (for the STAR collaboration), 
   J. Phys. G{\bf 34} (2007) S679. 

\bibitem{ridge2} 
   M.P. McCumber (for the PHENIX Collaboration), 
   J. Phys. G{\bf 35} (2007) 104081. 

\bibitem{dhump1} 
   M.J. Horner (for the STAR Collaboration), 
   J. Phys. G{\bf 34} (2007) S995.    

\bibitem{phobos} 
   E. Wenger (for the PHOBOS Collaboration), 
   J. Phys. G{\bf 35} (2008) 104080. 

\bibitem{phoboss} 
   B. Alver et al. (PHOBOS Collaboration), 
   Phys. Rev. Lett. {\bf 104} (2010) 062301. 

\bibitem{Feng:2008an}
  A.~Feng  [for the STAR Collaboration],
  %``Reaction Plane Dependent Away-side Modification and Near-side Ridge in
  %Au+Au Collisions,''
  J.\ Phys.\ G {\bf 35}, 104082 (2008)
  [arXiv:0807.4606 [nucl-ex]].
  %%CITATION = JPHGB,G35,104082;%%

\bibitem{Drescher:2000ec}
  H.~J.~Drescher, S.~Ostapchenko, T.~Pierog and K.~Werner,
  %``Initial condition for QGP evolution from NEXUS,''
  Phys.\ Rev.\  C {\bf 65}, 054902 (2002)
  [arXiv:hep-ph/0011219].
  %%CITATION = PHRVA,C65,054902;%%

\bibitem{Hama:2004rr}
  Y.~Hama, T.~Kodama and O.~.~J.~Socolowski,
  %``Topics on hydrodynamic model of nucleus nucleus collisions,''
  Braz.\ J.\ Phys.\  {\bf 35}, 24 (2005)
  [arXiv:hep-ph/0407264].
  %%CITATION = BJPHE,35,24;%%

\bibitem{hama:ismd2010}
  Y.Hama, R.P.G. Andrade, F. Grassi and W-L. Qian,
  in this proceedings.

\end{thebibliography}
\end{document}